\documentclass[acmsmall]{acmart}

\AtBeginDocument{%
  \providecommand\BibTeX{{%
    \normalfont B\kern-0.5em{\scshape i\kern-0.25em b}\kern-0.8em\TeX}}}

\usepackage{tikz}
\newcommand\encircle[1]{%
  \tikz[baseline=(X.base)] 
    \node (X) [draw, shape=circle, inner sep=0, fill=black, text=white] {\strut #1};%
    }

\usepackage{hyperref}

\setcopyright{rightsretained}

\acmConference[Group '22]{The ACM international conference on Supporting Group Work}{January 23–26, 2022}{Sanibel Island, Florida, USA}

\usepackage{ulem}
\usepackage{booktabs}
\usepackage{graphicx}
\usepackage{xcolor}
\usepackage{tabularx}

\definecolor{ourorange}{RGB}{252, 141, 98}
\definecolor{ourblue}{RGB}{141, 160, 203}
\definecolor{ourgreen}{RGB}{102, 194, 165}

\begin{document}

\title[Socio-Technical Contextualization of ML Interpretability]{Explanation Strategies as an Empirical-Analytical Lens for Socio-Technical Contextualization of Machine Learning Interpretability}

\author{Jesse Josua Benjamin}
\affiliation{%
  \institution{Department of Philosophy, University of Twente}
  \country{Netherlands}}
\affiliation{%
  \institution{Human-Centered Computing, Freie Universit\"at Berlin}
  \country{Germany}}
\author{Christoph Kinkeldey}
\affiliation{%
  \institution{Human-Centered Computing \\ Freie Universit\"at Berlin}
  \country{Germany}}
\author{Claudia M\"uller-Birn}
\affiliation{%
  \institution{Human-Centered Computing \\ Freie Universit\"at Berlin}
  \country{Germany}}
\author{Tim Korjakow}
\affiliation{%
  \institution{Human-Centered Computing \\ Freie Universit\"at Berlin}
  \country{Germany}}
\author{Eva-Maria Herbst}
\affiliation{%
  \institution{Human-Centered Computing \\ Freie Universit\"at Berlin}
  \country{Germany}}
\renewcommand{\shortauthors}{Benjamin et al.}

\begin{abstract}
During a research project in which we developed a machine learning (ML) driven visualization system for non-ML experts, we reflected on interpretability research in ML, computer-supported collaborative work and human-computer interaction. We found that while there are manifold technical approaches, these often focus on ML experts and are evaluated in decontextualized empirical studies. We hypothesized that participatory design research may support the understanding of stakeholders' situated sense-making in our project, yet, found guidance regarding ML interpretability inexhaustive. Building on philosophy of technology, we formulated explanation strategies as an empirical-analytical lens explicating how technical explanations mediate the contextual preferences concerning people's interpretations. In this paper, we contribute a report of our proof-of-concept use of explanation strategies to analyze a co-design workshop with non-ML experts, methodological implications for participatory design research, design implications for explanations for non-ML experts and suggest further investigation of technological mediation theories in the ML interpretability space.
\end{abstract}

%
%
%
%
%



\maketitle

\section{Introduction}

More than halfway into a three-year research project in which we, a group of human-computer interaction (HCI) researchers, developed a visualization system for research project data at a major natural history institution, we found ourselves at an unexpected impasse. Up to this point, we had followed a typical human-centered design methodology concerning socio-technical systems: we hosted data modelling workshops with institution stakeholders, conducted contextual inquiries, designed paper prototypes for potential visualizations and held semi-structured interviews with institution employees. Based especially on the latter, we concluded that a machine learning (ML) visualization system based on natural language processing (NLP) was predominantly called for. We hypothesized that clustering data based on thematic similarity would help to break down knowledge silos and tackle a lack of interdepartmental cooperation -- which were the primary issues our project was intended to overcome. However, when we presented the first prototype of our new ML-driven visualization system, we found that institution stakeholders struggled to interpret it. The issue, however, was not that stakeholders could not interpret it `correctly' in a technical sense. Instead, we found that interpretations, on the one hand, tended to perpetuate particular notions about the institutional context and, on the other hand, took the visualization ``as things really are'' rather than a data-driven proposal on possible opportunities for cooperation and knowledge exchange.
Subsequently, we turned to the then still developing discourse in HCI, ML and computer supported cooperative work (CSCW) on issues of ML interpretability and explainable artificial intelligence (XAI). Facilitating the interpretability of results generated by ML systems\footnote{ML systems are technological sub-systems that ``automatically improve from experience''~\cite{mitchell_machine_1997} by cyclically deploying algorithmic components for preprocessing data sets, inferring patterns from the data, and applying the resulting model for decision-making on new data.} has become one of the key challenges in HCI, CSCW and ML research. Indeed, we found a large corpus of research regarding the use of explanation methods (algorithmic techniques which extract data from ML systems, e.g. feature importance) to generate user-facing explanations of ML decision-making processes. However, studies of interpretability predominantly consider only an ML expert or data scientist perspective, are often model-centric, or provide a mathematical assessment of the model results ~\cite{abdul_trends_2018, carvalho2019MLinterpretability}. However, studies have also shown that notwithstanding such expertise, explanations can be over-accepted even if they are randomized or completely disconnected from the actual algorithms implemented (cf.~\cite{springer_dice_2017,kaur_interpreting_2020}). Additionally, more recent work emphasizes a clear urgency for advancing our understanding of non-ML experts' (e.g., subject matter experts, lay users) interpretability needs in a specific context of use (cf.~\cite{yang_investigating_2018,yang_mapping_2018, EhsanRiedl2020:HumanCenteredXAI, Shneiderman_2020:humancenteredAI}).

However, we could not discern guidance on how explanation methods and the explanations they generate map to actual practices of interpretation in socio-technical contexts. In order to operationalize this challenge, we formulated a research question to address the associated gap in related work: \textit{How do non-ML experts form interpretations when they interact with explanations of ML systems in actual contexts of use?} We subsequently reflected on recent proposals that participatory design research is generally well positioned to investigate ML interpretability for non-ML experts due to its openness towards situated sense-making and contextual socio-material practices (e.g.,~\cite{bannon_reimagining_2018}). We, therefore, planned to conduct a co-design workshop in which stakeholders engage potential explanations of the ML-driven visualization system. However, we also observed that more specific methodological steps and analytical lenses are needed which connect the \textit{technical} dimension of explanation methods to the ways in which specific explanations lead to particular interpretations by non-ML experts. We found a first indication in recent adaptations of theories of technological mediation in HCI research (e.g.,~\cite{verbeek_beyond_2015,hauser_annotated_2018,frauenberger_entanglement_2019}), where technological artefacts are seen to mediate how people perceive, reason and act in the world. We found a key conceptual framing in Hubig's related philosophy of technology: \textit{explanation strategies,} which indicate how technological artefacts lead people to infer specific, contextually meaningful sense-making practices~\cite{hubig_kunst_2015}. In our co-design workshop, we investigated our hypothesis that \textit{technologically mediated explanation strategies} offer an empirical-analytical lens to understand more clearly how particular \textit{explanations} (e.g., data visualizations), based on specific \textit{explanation methods} (e.g., feature importance), mediate non-ML experts' interpretations of ML systems.

As its main contribution, this paper explores how our framing of Hubig's explanation strategies can contribute to unfolding the contextual sense-making practices of non-ML experts through application in a co-design workshop. In the following, we first outline the current state of ML and CSCW/HCI interpretability research, discuss the promise of participatory design research to address shortcomings in the status quo and propose that technologically mediated explanation strategies offer a heretofore missing empirical-analytical lens in this space. We then detail our proof-of-concept application of explanation strategies in the context of a co-design workshop on explanation methods with subject matter experts. We outline the insights that the explanation strategies discerned offer for our use case and then elaborate in a discussion on the implications of our work for participatory design methods, the design of explanations and the use of technological mediation theories in the ML interpretability space.

We contribute the following with our research:
\begin{enumerate}
    \item proposal of an empirical-analytical lens for ML interpretability for non-ML experts in the form of \textit{technologically mediated explanation strategies};
    \item proof-of-concept detail on the analytical procedure and insights derived from applying this lens in the context of a co-design workshop; and
    \item provision of methodological, design and theoretical implications for future research and a proposal for using technological mediation theories in the field of ML interpretability.
\end{enumerate}

\section{Background}

In this section, we consider the challenge of ML interpretability for non-ML experts (e.g., subject matter experts, lay users) from a socio-technical perspective concerning three aspects. We firstly highlight how situated sense-making by non-ML experts of real-world ML-driven systems is rarely studied. Secondly, we observe that participatory design research is well suited as an approach towards this challenge. However, we find that an actionable concept for mapping technical explanations to situated sense-making is missing. This missing guidance is why we, thirdly, propose a bridging concept in the form of technologically mediated explanation strategies. This concept, we argue, provides an empirical-analytical lens for exploring situated sense-making in the context of explanations.

\subsection{Related Work from Interpretability Research}\label{ssec:related_work} 

The term ``interpretability'' is used widely in HCI/CSCW and ML research, although it encompasses numerous definitions~\cite{miller_explainable_2017}. Interpretability generally refers to techniques that make the behavior and predictions of ML systems understandable to humans. Interpretability techniques can be organized into data interpretability (e.g., exploratory data analysis techniques), intrinsic interpretability (e.g., lasso regression technique), and post-hoc interpretability~\cite{carvalho2019MLinterpretability}. In the context of our research, we focus on post-hoc interpretability, which is realized by XAI algorithms (e.g., LIME, SHAP). These algorithms can  be further specified as explanation methods (e.g., local methods, global methods)~\cite{liao_questioning_2020}, and these methods provide explanations (e.g., textual, visual). 

It is essential to highlight the difference between explainability and interpretability. An explanation, as delineated above, is a ``product'' of a specific XAI algorithm. According to the pragmatic theory of explanations, an explanation should not only provide a mathematically correct answer but also make the audience understand the reasoning behind a decision or prediction made by an ML model~\cite{kim2016criticisminterpretability}. Thus, to ensure interpretability, we need to understand explanations as a ``process''~\cite{miller_explainable_2017} which comprises a cognitive and a social dimension. The former relates to a process of abductive reasoning from which the actual explanation is derived. The latter relates to social interaction between the explainer, i.e., the ML-driven system, and the explainee, i.e., the human. This social dimension refers to the concept of situatedness, which highlights the need for an understanding of the ``local, situated practices of users'' when designing such interactions~\cite{harrison_making_2011}. 

However, when reviewing existing studies on ML interpretability research, we find that, on the one hand, many studies consider predominantly expert audiences, specifically people with formal ML education (e.g.,~\cite{cai_human-centered_2019,hohman_gamut:_2019,liao_questioning_2020,madaio_co-designing_2020, Hong_Hullman_Bertini_2020}). On the other hand, when studies consider lay users, their evaluation relies on crowdsourcing platforms (e.g.,~\cite{ribeiro_why_2016,cheng_explaining_2019,yu_keeping_2020}), which decontextualize interpretability. Thus, existing studies are not easily transferable to real-world contexts with lay users or subject-matter experts. Firstly, ML experts may prioritize values (e.g., performance metrics, such as accuracy, precision, recall) in ML-driven systems differently than stakeholders with no ML expertise or deeper knowledge in this area (they might, e.g., prioritize contextual notions such as meaningful ties among colleagues or infrastructures), or the abstract values, such as fairness, themselves may not provide a basis for interpretation (e.g.,~\cite{benjamin_materializing_2019,madaio_co-designing_2020}). Secondly, ML experts' prior decisions on suitable explanation methods shape what may be considered `valid' interpretations of ML-driven systems, possibly precluding those that would be contextually more meaningful for lay users~\cite{cramer_not-so-autonomous_2017}. Saha et al., for example, have observed that lay users struggle to understand fairness explanations of hiring predictions designed by ML experts as terms such as false negatives were hard to understand~\cite{saha_measuring_2020}. 
More contextually oriented proposals for human-centered ML seek to reframe existing ML workflows based on situated human work practices~\cite{gillies_human-centred_2016,fiebrink_introduction_2018, EhsanRiedl2020:HumanCenteredXAI}. However, Chancellor et al. found that current human-centered ML approaches are unclear on how this goal can be reached methodologically, leading to diverse and sometimes contradictory notions on how to give agency to stakeholders~\cite{chancellor_who_2019}. As a result, the proposals of human-centered ML have not yet broadly caught on regarding ML interpretability for non-ML experts. 
Lastly, studies have also found that explanation methods can lead to an \textit{over-acceptance} of ML outputs, both for ML experts and lay users~\cite{kaur_interpreting_2020, springer_dice_2017}. Specifically, Yang et al. have found that non-ML experts tend to overestimate the significance of performance metrics~\cite{yang_grounding_2018}. Such studies suggest that more research should be done to understand the `hows and whys' of interpretation, particularly when lay users interact with explanations of ML-driven systems.

To summarize, research on explanations provided by ML-driven systems in real-world contexts with non-ML experts remains rare~\cite{abdul_trends_2018,sendak_human_2020}. Existing studies are often predominantly expert-oriented and rely on decontextualized settings, both in terms of participants (crowdworkers) and technologies (specific data sets, selected XAI algorithms). This leaves the domain expertise of subject matter experts often unattended and not reflected in the explanations provided. Furthermore, a clear guidance on how to design explanations for non-ML experts in their specific contexts of use is still missing. In the following section, we discuss how participatory design research provides a first direction for studying non-ML experts' interpretations of explanations provided.

\subsection{Promise of Participatory Design Research for Designing ML-driven Systems}
\label{ssec:pdML}

Participatory design methods have been argued to be particularly suited for studying the complexities of situated sense-making with ML-driven systems, as they are concerned with the multiplicity of world-views that stakeholders exhibit both implicitly and explicitly (e.g., ~\cite{bannon_reimagining_2018,loi_co-designing_2019,wolf_conceptualizing_2019, yang_re-examining_2020, Hong_Hullman_Bertini_2020}). Such proposals are based on the general methodological finding that participatory design research methods have been particularly effective for understanding how novel technologies are engaged in stakeholders' situated sense-making (e.g.,~\cite{ehn_participation_2008}).

In a notable recent example, M\o ller and colleagues employed co-design methods with multidisciplinary stakeholders to disclose and problematize values in the design of a civic job placement scheme ~\cite{holten_moller_shifting_2020}. While this showcases the suitability of participatory design methods in general for current socio-technical issues associated with ML, it does not specifically address the design of explanations and its relation to contextual sense-making practices. 
A recent example regarding ML interpretability is Dove and Fayard's work on using co-design methods to design meta-concepts for AI~\cite{dove_monsters_2020}. However, their method and results are not directly related to an \textit{actual} ML deployment, and transferability to real-world use cases remains unclear. Current research from the medical domain where co-design workshops are used to understand interpretability requirements of medical professionals is more closely related to real-world systems (e.g.,~\cite{wang_designing_2019, xie_chexplain_2020}). However, the analyses of workshops were specifically oriented to the cognitive processes of error-correction rather than the contextual reasons why ML outputs are understood in specific ways, with the authors calling for more fundamental research on this issue. In addition to conducting co-design workshops, ``explainability scenarios''~\cite{wolf_explainability_2019,andres_scenario-based_2020} provide a promising approach for reflecting technical explanation issues within contextual interpretability settings. 
This idea has been adapted by Ehsan et al., who used scenarios in an interview study to explore the role of the socio-organizational context in ML-based applications~\cite{Ehsanetal2021:Expanding}. Similarly, Liao et al. and Wang et al. show how to develop contextually sensitive questions~\cite{liao_questioning_2020,wang_designing_2019} for various stakeholders by considering explanation needs and how these questions can be used for a question-driven design process~\cite{Liao2021:Questiondesignprocess}.

Even though this research is valuable for understanding the role of context for interpretability and provides first methodological guidance of mapping XAI algorithms to specific questions (e.g., what if, how), an understanding of ``How is it that stakeholders form interpretations when they interact with explanations in actual contexts of use?'' is not exhaustively elaborated. Broadly speaking, it is unclear what concepts and methods can link participatory approaches to situated sense-making with ML's ``statistical intelligence''~\cite{dove_ux_2017}. Whereas participatory design methods, such as co-design, are highly promising to fill the gap of studies with non-ML experts in interpretability research, it seems that HCI and CSCW lack an empirical-analytical approach for exploring the relationship between specific explanations and their situated interpretation. Since we experienced this challenge in our research, we have adapted an approach from philosophy of technology to fill this gap. We detail this approach next and afterward describe the real-world situation that led to this approach in more depth .

\subsection{Developing an Empirical-Analytical Approach to Interpretability}\label{ssec:expl}

Researchers (e.g.,~\cite{miller_explainable_2017,miller_explanation_2017}) propose that ML interpretability research is increasingly making use of theoretical frameworks from the humanities and the social sciences. Wang et al.~\cite{wang_designing_2019}, for example, have developed a conceptual framework for mapping human reasoning onto explanation methods. As we are looking for an empirical-analytical dimension to understand how people take up ML explanations (generated from XAI algorithms) within their everyday context, we note Miller's foregrounding of \textit{abductive reasoning}~\cite{miller_explanation_2017}, which relates to the cognitive dimension of understanding explanations as a process.\footnote{As opposed to deduction from axioms, or induction from observed phenomena, abduction refers to the selection of the most likely explanation for a given phenomenon from other (possibly unknown) explanations.} He proposes that this type of reasoning is most `natural,' in the sense that it characterizes everyday situations (i.e., unlike laboratory conditions or scientific inquiry). However, even though abductive reasoning seems promising for approaching explanations as a process, Miller does not yet offer a precise grasp on how it may be applied in specific interpretability contexts. 

The influx of theoretical scaffolding from the humanities for understanding contextual sense-making practices has a long tradition in HCI and CSCW research, particularly following the introduction to the field of ethnomethodology in the works of scholars such as Suchman, Dourish or Crabtree (e.g.~\cite{suchman_plans_1990,dourish_technomethodology:_1998,crabtree_ethnomethodologically_2000}), respectively. Broadly concerning the practices of meaning-making people achieve in day-to-day activities, i.e. ethnomethods, this phenomenology-inspired sociological approach has long shaped contextually oriented HCI and CSCW research, for example, in research on how end users integrate technologies in workplace settings (e.g., ~\cite{dourish_appropriation_2003}). The need for a specific empirical-analytical approach to ML interpretability contexts with actual systems and actual people, therefore, can be seen as a continuation of this research agenda. Given the methodological consequences of ethnomethodology in HCI and CSCW research, such as the use of grounded theory (cf.~\cite{muller_grounded_2010}), there are also promising overlaps regarding the analysis of abductive reasoning (e.g.,~\cite{patokorpi_what_2009}). However, we argue that a more specific lens is required given the particular challenges of ML interpretability concerning technical origins (i.e., explanation methods) and the things that \textit{can} be appropriated (i.e., explanations) by people.

An empirical-analytical dimension for studying people's abductive reasoning in relation to technological artefacts can be found in the work of philosopher of technology Hubig. Similar to Miller, Hubig argues that everyday human sense-making with and via technologies occurs predominantly in the mode of abductive reasoning~\cite{hubig_kunst_2015,richter_space_2020}. Hubig posits that this is an effect of \textit{technological mediation}: facts and statements mediated by technological objects are not deductions or inductions from `objective' phenomena but rather abducted from technologically generated phenomena. The point here is that technologies `themselves' have no inherent notion of meaning, therefore, humans must derive heuristic assumptions (consciously or not) from abductive reasoning processes to come to conclusions about technological outputs. Hubig proposes \textit{explanation strategies} to account for the type of abductive reasoning that stems from a rich, socio-historical background encountered in specific contexts -- the kind of reasoning that we become accustomed to and which we rely on to navigate day-to-day concerns.\footnote{Hubig furthermore distinguishes between four fields (perceptive, semantic, causal, presuppositional) of abduction, which apply across four types of abductive reasoning (abductive induction, hypostatic abstraction, explanation abduction, explanation strategy abduction). We focus exclusively on the last type as a socio-technical category within the scope of our paper.} Accordingly, this perspective emphasizes the role of human sense-making and context, and highlights that a focus on explanations as a product (e.g., of an XAI algorithm) is not sufficient.

\begin{figure}[tb]
  \centering
  \includegraphics[width=0.8\columnwidth]{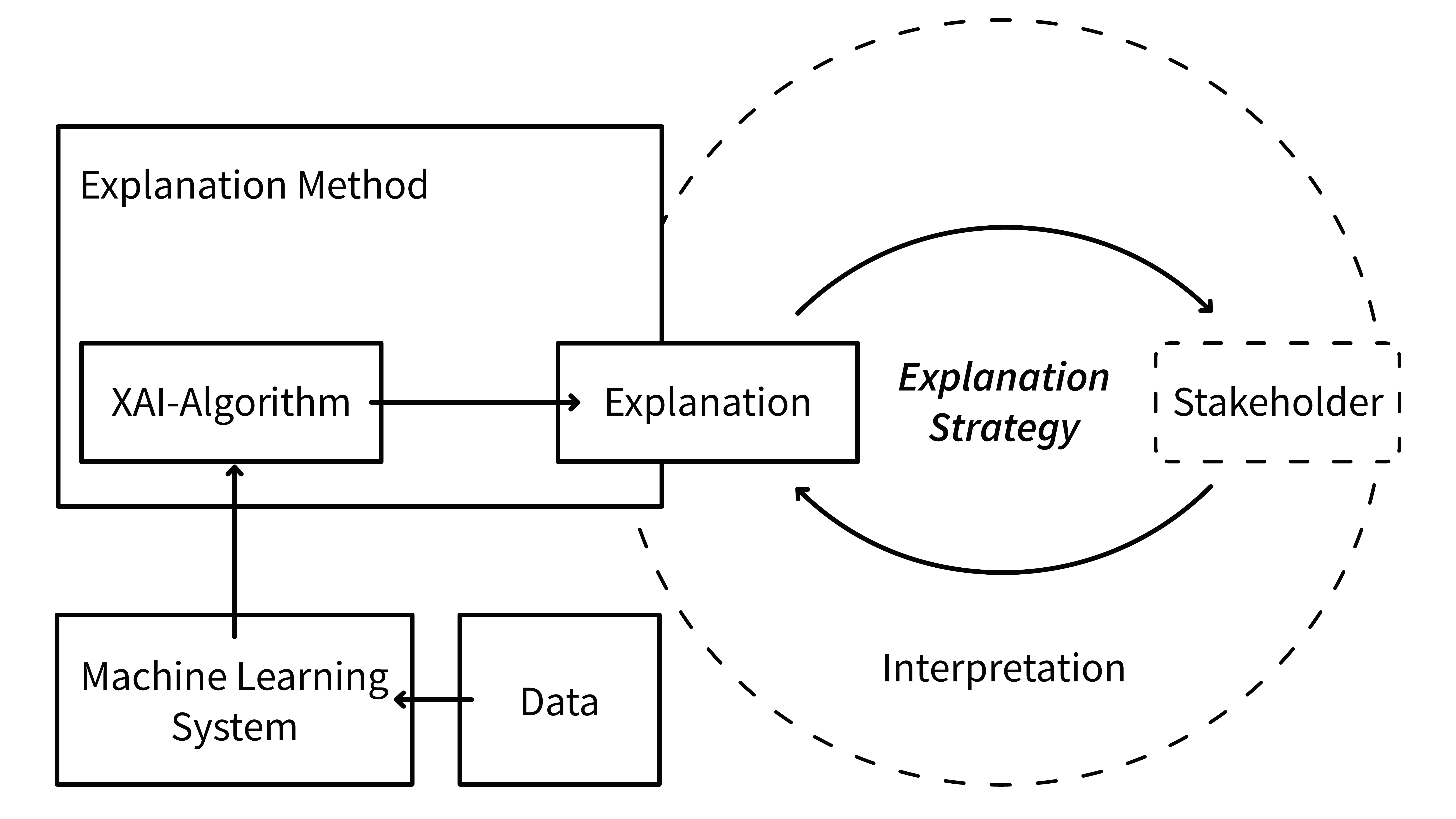}
  \caption{\label{fig:diagramm}%
           A diagram explaining our hypothesis on the relationship of algorithmic explanation methods, explanations, and explanation strategies. An explanation method applies explainable artificial intelligence (XAI) algorithms to generate an explanation relating to a specific processing step in an ML system. A contextually informed and technologically mediated explanation strategy manifests as a concrete interpretation in the interaction between explanation and stakeholder.}
    \Description{A component diagram of our proposed hypothesis on the relationship of algorithmic explanation methods, explanations, and explanation strategies. An explanation method applies explainable artificial intelligence (XAI) algorithms to generate an explanation relating to a specific processing step in an ML system. A contextually informed and technologically mediated explanation strategy manifests as a concrete interpretation in the interaction between explanation and stakeholder.}
\end{figure}

In CSCW and HCI design research, philosophical frameworks on technological mediation, such as post-phenomenology or actor-network theory have become widely used, grounding what Frauenberger has recently described as ``entanglement HCI''~\cite{frauenberger_entanglement_2019}. We posit, therefore, that Hubig's \textit{explanation strategies}, as they build on similar premises, may provide a suitable level for HCI researchers to analyze how technical explanations mediate specific practices of situated sense-making. 
We consider explanation strategies as a promising \textit{empirical} dimension, insofar as it can be used to study concrete deployment contexts and how people reason in them, and an \textit{analytical} dimension because explanation strategies can differentiate between the heuristic assumptions that underlie such reasoning. 
Therefore, we hypothesize that studying how the explanation strategies of non-ML experts are mediated by specific explanations can provide researchers with an understanding of how the socio-organizational context, human reasoning and the explanations provided lead to specific interpretations (cf.~\autoref{fig:diagramm}).

\begin{table}[tbh]
\begin{tabularx}{\textwidth}{@{}llll@{}}
\toprule
\textit{Explanation Strategy} & Paradigmatic & Conceptual & Presuppositional \\ \midrule
\textit{Scope} & Patterns of experience & Generative & Frames of reference \\
\textit{Example} & ``We always do & ``Maybe we can see & ``We have always \\
 & x this way.'' & x through y?'' & been x.'' \\ \bottomrule
\end{tabularx}
\caption{Explanation strategies which help one to understand how non-ML experts make sense of explanations in specific contexts (adopted from Hubig~\cite{hubig_kunst_2015}).}
\Description{A table providing an overview of the explanation strategies we hypothesize can help researchers understand how non-technical stakeholders in specific contexts make sense of technological artefacts. Paradigmatic abductions manifest in explanation strategies that pursue interpretations based on specific, practical examples (e.g., "We should see x this way."). Conceptual abductions manifest in explanation strategies that pursue new lines of interpretation (e.g., "Maybe we can see x through y?"). Presuppositional abductions manifest in explanation strategies that rely on general frames of reference for interpretations (e.g., "We have always been x.").}
\label{explstrat}
\end{table}
\vspace*{-5mm}

As we will detail more fully below, we applied Hubig's explanation strategies in our own use case to understand how stakeholders made sense of the explanations provided. We, furthermore, built on Hubig to formulate three types of explanation strategies which we discerned directly from people's statements and actions (as described in \autoref{sec:analysis}) to systematize our approach. 
Given specific prompts, \textit{paradigmatic} explanation strategies indicate how stakeholders integrate phenomena into previous practical experience (e.g., ``We have always done it this way before, and that is how \textit{x} can be done as well.''). Put differently, such explanation strategies are in play when events call for (implicit or explicit) interpretation following patterns of prior experiences. 
On a more general level, \textit{presuppositional} explanation strategies address the prior, often firm beliefs that stakeholders have about their context (e.g., ``As a technical department, we don't deal with social consequences.''). Technological mediation of explanation strategies in this field, then, concerns how explanations are seen or made to fit into stakeholders' existing presuppositions about `their' world, such as values and norms regarding context and technologies. Presuppositions in an HCI and CSCW context may be further understood as ``professional vision''~\cite{goodwin_professional_1994} or ``organizational acculturation''~\cite{reis_louis_acculturation_1990}: existing frames of reference and learned practices in socio-technical contexts.
We, furthermore, distinguish \textit{conceptual} explanation strategies in between these specific and general strategies. Such abductions occur when a novel position on a matter at hand is taken up (e.g., ``Maybe we can see \textit{x} through \textit{y}?''). Conceptual explanation strategies, therefore, may indicate when technologies mediate perspectives that lead to a generative reflection on the significance of technical outputs in light of their contextual deployment (e.g., applying new concepts to existing matters).

Taken together, these three types of explanation strategies address how technological artefacts, in our use case explanations, co-constitute~\cite{verbeek_what_2006} how people reason \textit{practically} (i.e., paradigmatic), \textit{generally} (i.e., presuppositional) and \textit{generatively} (i.e., conceptual) in a context of use (cf.~\autoref{explstrat}). 
In the following, we showcase how we applied this empirical-analytical dimension for ML interpretability in a co-design workshop. Our approach, we argue, is a promising first step to fill the gap of analyzing human sense-making of explanations in a concrete context of use.

\section{Surfacing Explanation Strategies in a Co-Design Workshop}

In the following, we describe the specific use case from which the need for a better understanding of people's sense-making processes in the context of ML explanations provided emerged. Our own experience motivated us to conduct a co-design workshop. In this context, we discovered and built on Hubig's work to investigate whether explanation strategies can provide a suitable empirical-analytical dimension for studying interpretability in situated sense-making practices. Thus, we outline how we prepared and deployed this workshop and analyzed the results.

\subsection{Interpretability Concerns in an ML-driven Visualization System Use Case}
\label{ssec:usecase}

As indicated in our introduction, we developed an ML-driven visualization system for a natural history research institution to address a self-reported lack of knowledge exchange between subject matter experts. The visualization system is intended to facilitate cross-domain cooperation among the institution's stakeholders, who are from various domains of expertise and organized in distinct research departments. A pre-study conducted with these stakeholders\footnote{The semi-structured interview study is described in detail in a technical report, see ~\cite{benjamin_understanding_2019}.} resulted in a visualization design which displays thematic similarities in research projects.\footnote{The data and source code of the ML-driven visualization system is available in an open source GitHub repository \url{https://github.com/FUB-HCC/IKON-backend}, accessed 09/24/2021.} We hypothesized that such a visualization would encourage the discovery of shared practices and, therefore, foster engagement in knowledge exchange by our stakeholders. 
%
%
%
The visualization we developed displayed research projects based on their thematic similarity. In order to create this visualization, we prepared and transformed data on existing research projects at the institution in a NLP pipeline. The latter comprises four components that correspond to typical processing steps: preprocessing of the data, semantic analysis, clustering and dimensionality reduction for visualization. The output, at this stage, is a clustered scatterplot of the data, i.e., the research projects separated according to the similarity inferred from semantic analysis. 

However, when presenting this first visualization prototype to our stakeholders, they had difficulties to make sense of the results. The cluster visualization was, for example, taken to ``show things how they really are'' instead of seeing \textit{possible} similarities between research projects. This missing basis for stakeholder interpretations was troublesome as our use of `traditional' human-centered HCI methods had neither brought this specific problem to the surface nor suggested a way forward. The difficulty for us was \textit{not} whether our stakeholders could interpret our visualization in one specific way, but, instead, in which ways our stakeholders would be able to reflect on and integrate the `ML model's view' on the data into their own interpretations of our NLP pipeline's output. We assumed that, as suggested by the related work, including explanations in our visualization design could address these concerns and facilitate interpretation (cf.~\autoref{ssec:related_work}). However, we could not find guidance on how potentially applicable explanations would actually map to a stakeholder's situated sense-making practices. It is from this set of circumstances that we began to reflect more in-depth on suitable explanation methods for our visualization and first started to discuss whether explanation strategies might help us to understand more deeply how human sense-making, technical explanations and context become integrated. Thus, we decided to develop a co-design workshop with the aim of understanding in which ways the explanations provided mediate non-ML stakeholders' contextual explanation strategies.

\subsection{Development of our Co-Design Workshop}
\label{ssec:PIMeth}

In this section, we detail our rationale for the specific workshop we developed, the artefacts we designed for it, the explanation methods used and the developed workshop procedure.

\subsubsection{Rationale for a Co-Design Workshop}
Given the interpretability concerns above and a lack of understanding regarding how exactly these concerns surface in the application context, we needed to find a way that would bring our stakeholders together with possible explanations. We assumed that a co-design workshop in which stakeholders would engage with possible explanations would allow us to understand why specific interpretability concerns emerge in our use case context more precisely. We, therefore, base our co-design workshop on the understanding of participatory design by Schuler and Namioka~\cite{schuler_participatory_1993} as discussed by Halskov and Hansen~\cite{halskov_diversity_2015}: what is \textit{co-designed} in our workshop is not so much a concrete user interface but rather a contextually refined \textit{understanding} of the latter by considering our stakeholders as ``experts in their worklife'' whose ``perception of technology [is] important.'' The immediate, internal goal of the co-design workshop was to improve our ML-driven visualization system. However, the co-design workshop both provoked and allowed us to test how far the empirical-analytical dimension of explanation strategies could be extended towards participatory methods more generally. We thereby sought to generate ``intermediate-level knowledge''~\cite{hook_strong_2012} from the design, deployment and analysis of our co-design workshop that would offer pointers to future research in this domain.

\begin{figure}[t!]
  \centering
  \includegraphics[width=0.9\linewidth]{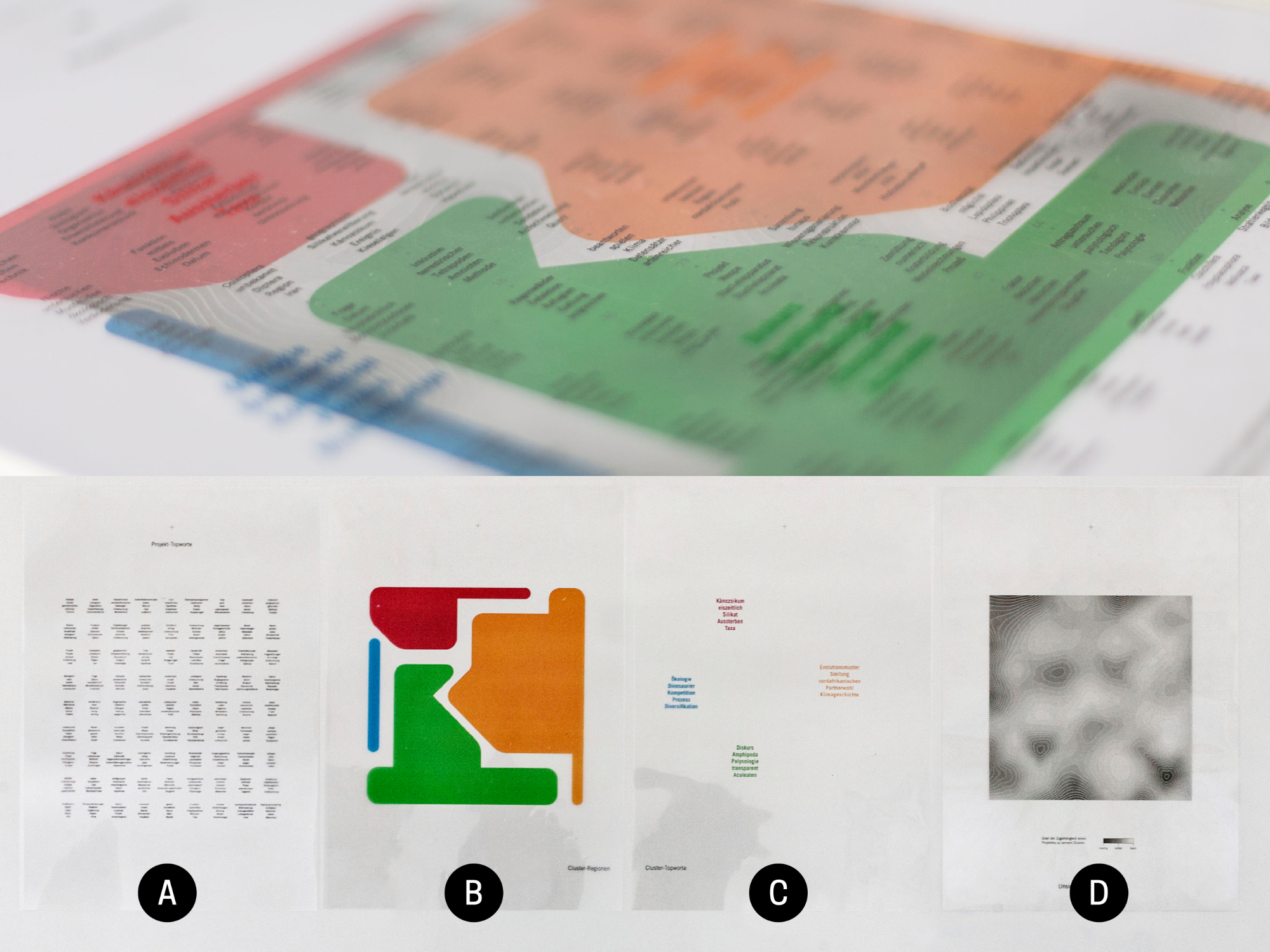}
  \caption{
        \textit{Top:} Overlay effect of transparencies; \textit{Bottom, left to right:} Transparencies -- project topwords, cluster regions, cluster topwords, uncertainty landscape.}
    \Description{A composite image, showing the effect of overlaying the designed transparencies on top, and individual transparencies from left to right on the bottom.}
    \label{fig:material}
\end{figure}

\subsubsection{Categorizing and Selecting Explanation Methods for the Use Case NLP Pipeline}

We built upon existing work on model interpretability by Hohman et al.~\cite{hohman_gamut:_2019} to select suitable explanation methods for our use case. Since we could not predict which particular scope of explanations was needed in our use case based on related work, we considered methods that explain the whole model and methods that focus on a single data instance: 

\begin{itemize}
    \item[] \smash{\encircle{\textbf{\texttt{A}}}} \textit{\textbf{`Project Topwords' --}} A local feature importance technique (e.g.,~\cite{green_generating_2009}) for individual predictions on the most influential topics for each research project showing groups of five words per project. Each word group was extracted from the semantic analysis component of our NLP pipeline, representing the five most significant words from the abstracts of the research project that have led the respective project to be positioned. When the transparency is put onto the project index, the project topwords appear underneath their respective grid point, i.e., project. We assumed that this transparency would mediate paradigmatic explanation strategies, indicating how our participants reason about how ``things ought to be done'' regarding research projects.
    \item[] \smash{\encircle{\textbf{\texttt{B}}}} \textit{\textbf{`Cluster Regions' --}} A global similarity technique (e.g.,~\cite{balachandran_interpretable_2012}) grouping projects by the clusters derived by the k-Means algorithm and displaying the clusters as concave colored polygons. We assumed that this transparency would mediate presuppositional explanation strategies of global patterns and, thereby, indicate the kind of `macro-structures' (e.g., values and norms regarding topics, research groups) that participants considered specifically relevant for their socio-material context.
    \item[] \smash{\encircle{\textbf{\texttt{C}}}} \textit{\textbf{`Cluster Topwords' --}} A  global feature importance technique (e.g.,~\cite{ish-horowicz_interpreting_2019}) providing the most influential topics in semantic analysis on the more general level of clusters instead of projects, cf. `Project Topwords' transparency. We assumed that this transparency would mediate presuppositional explanation strategies similar to the cluster regions but, more specifically, on the level of domain expertise.
    \item[] \smash{\encircle{\texttt{\textbf{D}}}} \textit{\textbf{`Uncertainty Landscape' --}} A regions of error technique (e.g.,~\cite{hohman_gamut:_2019}), describing how well projects are represented by their cluster and, thereby, offering a general perspective of the NLP pipeline output. It displays a grey scale `topography' that when placed on the project index, shows individual projects in `slopes,' `peaks' or `valleys.' This, indicated by a legend, suggested how well a project `fits' into its respective cluster according to the clustering component (k-Means). We assumed that this would mediate paradigmatic and presuppositional explanation strategies in which participants consciously weighed their own perspective on the socio-material context against the ML results, for example, by taking the overall clustering `with a pinch of salt.'
\end{itemize}

\subsubsection{Designing Workshop Artefacts from Technical Explanations}
\label{ssec:tangibles}

A major concern was the extent of participation that the explanations provided could actually foster. After all, our design of the outputs of explanation methods would shape the ways and means in which participants could pursue their explanation strategies~\cite{odom_research_2016}. We wanted to prevent `usability traps' that would lead to `user design' as opposed to participation~\cite{vines_configuring_2013}. Accordingly, we designed physical artefacts exclusively to avoid a focus on interface elements. Based on these design rationales, we engaged in an iterative design process.\footnote{An OSF download \url{https://osf.io/8tfzb/download} for sketches and details is available.} We decided to design possible explanations as transparencies that could be placed on a physical printout of our cluster visualization. We designed our artefacts to be representative of the ML-driven visualization itself and created (1) the \textit{project index} (a heavy card base featuring identification numbers for 81 research projects on a 9 x 9 grid), (2) the \textit{research book} (a ring-bound book composed of 81 project fact sheets (title, abstract and index number) in numerical order), (3) the \textit{task book} (a ring-bound booklet composed of eight tasks) and (4) the \textit{transparencies} (overlay transparencies that represented the individual explanations, cf.~\autoref{fig:material}). The design of workshop artefacts informed our development of a workshop procedure directly.

\subsubsection{Developing a Workshop Procedure}

Taking inspiration from Hsueh et al., we wanted to ensure our participants could shift between reactive and self-reflexive modes of engagement with our artefacts~\cite{hsueh_understanding_2019}. We, therefore, split the workshop into two parts each with two phases, respectively. 

\begin{figure}[tb]
  \centering
  \includegraphics[width=1\columnwidth]{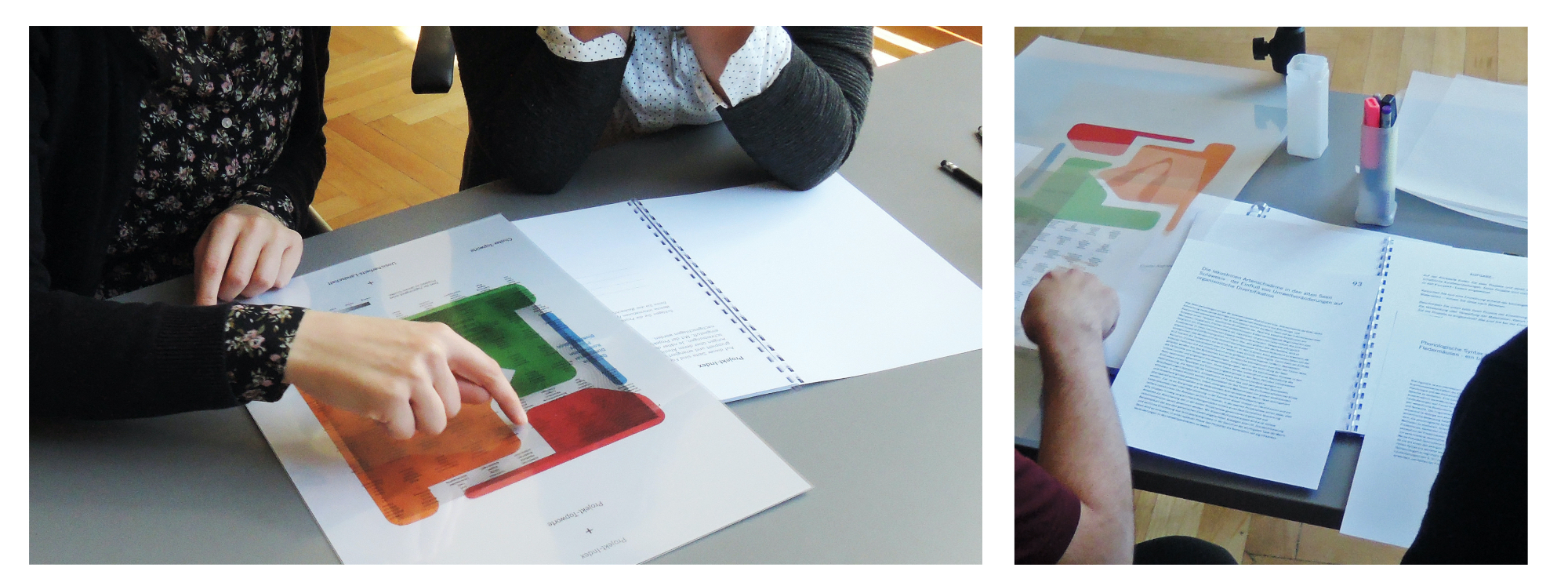}
  \caption{\label{fig:part1}%
           \textit{Left:} Participants discussing a task (phase I). \textit{Right:} Participants preparing to emulate the ML-based visualization (phase II).}
    \Description{ Photos from phases 1 and 2. On the left, participants are bent over a stack of transparencies, one of them pointing and the other pondering the assigned task in phase 1. On the right, in an over-the-shoulder shot, participants are preparing their data classification in phase 2, and have a fact sheet on the new data instances between them.}
\end{figure}

Participants engaged the workshop artefacts and the cluster visualization in the first, \textit{reactive} part.
\begin{itemize}
    \item \textit{Phase I --- Application:} After a short orientation, we provided each group with the project index and the explanations and asked our participants to work on the six tasks presented in the task book; a series of low-level, open-ended tasks typical for the visualization domain (e.g.,~\cite{amar_low-level_2005}). Each task introduced a new transparency use without predetermining valid solutions (cf.~\autoref{fig:part1}, left), such as ``Put the `project top words' transparency onto the project-index and locate projects 52 and 46. What can you find out about the projects?'' This phase was designed to learn about the explanation strategies that participants pursued with the respective explanation method and familiarize participants with the material at hand.
    \item \textit{Phase II --- Enactment:} In this phase, we provided the participants with a description (fact sheet, cf.~\autoref{fig:part1}, right) of two research projects that were not represented in the project index provided. The task was to assign the projects to a location in the project index by using any of the techniques. We hypothesized that if participants were to enact the ML processes of data transformation by way of our transparencies, they would reflect more closely on whether and which explanation methods actually mapped to their sense-making practices; which would allow us to infer contextual explanation strategies.
\end{itemize}

\begin{figure}[tb]
  \centering
  \includegraphics[width=1\columnwidth]{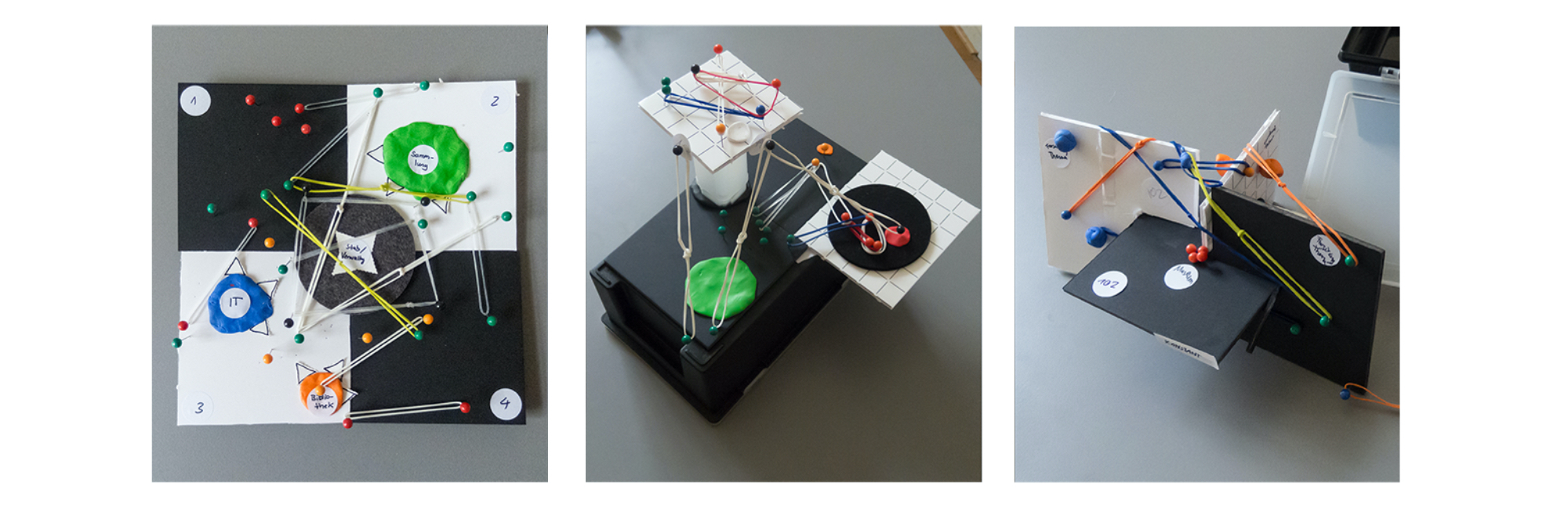}
  \caption{
           From left to right, tangibles built by groups A / B / C (phase III \& IV).}
    \Description{Images of the representations that participants constructed in phase 3. On the left, group A built a checkerboard-like flat representation, with black and white panes alternating to show strict departmental separations. There are strings connecting projects and infrastructures between departments. In the middle, group B built a towering construct, in which departments are distributed at various heights and distances, with interconnecting strings on top of a black box. On the right, group C constructed an abstract representation, with black and white panes intersecting which represents the temporally distributed relevance of research topics against the fixed institutional center.}
    \label{fig:part2}
\end{figure}

In the second, \textit{reflexive} part of the workshop, our participants made their own representations of the use case context.
\begin{itemize}
    \item \textit{Phase III --- Reification:} Participants were first asked to represent their own view of `research at your institution.' We assembled a design toolkit for this phase with prototyping material, for example, modelling clay, various surfaces, strings and pins. We assumed that, as our participants were primed by the prior reactive phases, this phase would explicate the contextual presuppositions which may have guided participant decision-making in the more task-oriented phases I and II.
    \item \textit{Phase IV --- Reflection:} In the final phase, groups exchanged the objects they had created in phase III (cf.~\autoref{fig:part2}). Mirroring phase II, we provided the participants with two further project fact sheets and asked them to integrate these into the other group's object in whatever way they saw fit, for example, using pins or stickers to indicate where a project may reasonably be located in the physical representation given. Thus, we hoped that the interpretation of another group's object would indicate differential presuppositions, which, in combination with the other phases, would offer further insights into existing presuppositional explanation strategies.
\end{itemize}

\subsection{Deployment of the Workshop}

Our workshop lasted three hours, with a short break between phases II and III.\footnote{We provide an OSF download \url{https://osf.io/bv3wc/download} for material on our workshop delivery.} 
We recruited subject matter experts among the staff of the natural history research institution and, finally, selected six participants (two female, four male; self-reported). All of our participants fulfilled different roles at the institution (e.g., research, exhibition design, knowledge repository development and science communication). The selection of the participants was motivated by the overall use case goal of supporting knowledge exchange and collaboration. We conducted the workshop in a meeting room at the institution.\footnote{The workshop was deployed prior to the Covid-19 global pandemic.} We used a camera to support our data collection by logging interactions with the workshop artefacts and among participants.


In the workshop, our participants were asked to form three groups of two in order to support the social dimension of the interpretation process (cf.~\cite{miller_explanation_2017}). A post-workshop questionnaire showed the diverse expertise of our participants and reflected both individual and institutional biographies: one participant had spent their entire ten-year career at the institution (P1), whereas another (P3) had only joined two years ago after fifteen years elsewhere. 
The three groups can be characterized by: A (P1 \& P2): prior acquaintance and a shared organizational background but with a disparity of expertise and research experience at the institution; B (P3 \& P4): little prior acquaintance and a shared strong research background in general but relatively little time spent at the institution and mixed expertise; and C (P5 \& P6): no prior acquaintance and equally non-research related positions at the institution and mixed expertise. Each group was accompanied by one researcher, who observed the group and took notes on discussions and bodily behavior.

\subsection{Analysis}
\label{sec:analysis}

Our analysis was developed and conducted by three attending researchers subsequent to the workshop.\footnote{All analysis data is provided in an OSF download: \url{https://osf.io/gcp4j/download}.} We collected more than nine hours of audiovisual footage, the participants' task books, the objects that were created, markings and notes on pages and the transparencies, photos and our own notes. We transcribed all group video footage individually, noting both dialogue between and statements made by participants. The data set consisted of the transcriptions (phases I--IV), the altered workshop artefacts (phases I--II) and images of the objects created by the participants (phases III--IV).
In the workshop, we expected participants to make ``contextual attributions''~\cite{miller_explanation_2017} which we could base on an interpretive thematic analysis~\cite{peterson_thematic_2017} of explanation strategies. In the first round of our thematic interpretive analysis, we discussed our impressions from our individually generated transcripts and formulated initial themes that we hypothesized would guide our analysis, which consisted of the explanation strategies assumed (cf.~\autoref{explstrat}) and participant behavior (e.g., searching, showing and solving). In a second round of analysis, we differentiated between paradigmatic, conceptual and presuppositional explanation strategies. We differentiated when participants chose specific examples in their interpretations (paradigmatic explanation strategy), based their interpretation on more general assumptions (presuppositional explanation strategy) or used explanations for unanticipated interpretations (conceptual explanation strategy). Through this differentiation of explanation strategies, we were able to describe more precisely which explanations strategies were at work regarding an explanation provided. Subsequently, each researcher then coded all transcripts individually. In our final round of analysis, we collaboratively derived our main findings by comparing and discussing the coded transcripts.


\begin{table}[tbh]
\begin{tabularx}{\textwidth}{lll}
\toprule
\textit{Explanation Strategy} & \textit{Use Case Context} & \textit{Explanations}\\ 
\midrule
\textit{Paradigmatic} 
& 
Contrast \& Utility  &
\encircle{\textbf{\texttt{A}}} Project Topwords  \encircle{\textbf{\texttt{C}}} Cluster Topwords \\ & (cf.~\autoref{sec:interpretation}) & \\
&
Justifying Hierarchies  &
\encircle{\textbf{\texttt{B}}} Cluster Regions \\ & (cf.~\autoref{sec:hierachies}) &
\\ 
\midrule
\textit{Conceptual}  
&
Algorithmic Perspective &
\encircle{\textbf{\texttt{D}}} Uncertainty Landscape \\
&
(cf.~\autoref{sec:attribution}) \\
&
Questioning Coherence &
\encircle{\textbf{\texttt{B}}} Cluster Regions \encircle{\textbf{\texttt{D}}} Uncertainty Landscape \\ 
&
(cf.~\autoref{sec:coherence})
&  \\
\\
\midrule
\textit{Presuppositional}
&
Organizational Hierarchy  &
Tangibles (cf.~\autoref{fig:part2}) \\
&
(cf.~\autoref{sec:hierarchy})
& \\
& Socio-material Relations \\
& (cf.~\autoref{sec:relations}) &
 \\
\bottomrule
\end{tabularx}
\caption{Overview of our findings organized into the type of explanation strategy (column 1), their respective manifestation in our use case context (column 2) and the related explanations/artefacts (column 3).}
\Description{Overview of our findings organized into the type of explanation strategy (column 1), their respective manifestation in our use case context (column 2) and the related explanations/artefacts (column 3).}
\label{tab:explstrat}
\end{table}
\vspace*{-8mm}

\section{Findings}

Our findings center on our hypothesis regarding how the explanations generated from explanation methods mediated specific explanation strategies for stakeholders in a socio-technical context (cf.~\autoref{fig:diagramm}). In the following, we differentiate between the explanation strategies defined above (cf.~\autoref{tab:explstrat}). To recap, we analyzed participant statements and actions for three kinds of explanation strategies. In our theoretical background, we had considered \textit{paradigmatic} and \textit{presuppositional} explanation strategies that would allow us to understand the interpretive backgrounds in our use case context; while \textit{conceptual} explanation strategies would indicate whether and how the ML-driven visualization system could potentially bring forth new participant perspectives on that context. We differentiate below between these findings accordingly.

\subsection{Paradigmatic Explanation Strategies}
The analytical focus in paradigmatic explanation strategies was how the explanations provided in phases I and II led participants to interpret specific issues related to existing experiences of `getting things done,' which indicate patterns of interpreting contextual matters. We identified two common themes from our participants.

\subsubsection{Utility and Contrast over Interpretation} 
\label{sec:interpretation}

We found throughout the task-oriented phase I that the explanations generated from feature importance techniques (\textit{project} and \textit{cluster topwords}) were generally mostly used to solve tasks. Participants actively compared the features to the words they had highlighted to quickly understand research project abstracts. P5, for example, noted that a project would fit its cluster as ``there are two words fitting together, <points to red cluster top words> 'glacial' and 'extinction'.'' On an initial reading, we could, therefore, see that feature importance techniques were indeed used to make sense of the project index. However, the explicit statements by participants also indicated that the explanation strategies mediated by this technique were of a utilitarian manner as opposed to reflections on how the cluster visualization of data mapped to the socio-material context. P2, for instance, compared features unfavorably to ``our own topwords [highlighted in research project abstracts],'' before dismissing concerns and solving the task at hand anyway. Furthermore, on the level of cluster topwords, we also found indications that sense-making statements bent towards contrasts between rather than synthesis of explanations and contextual knowledge. P3, for example, argued that `zoology' should not surface in two clusters as it is ``too large of a topic [at the institution].'' As a department which actually exists, `zoology' has a specific contextual meaning which \textit{supersedes its presentation as a data label}. This indicates that feature importance techniques in our use case context could cause the discussed effect of `over-acceptance' of explanations, on the one hand, or lead to a rejection of the visualization due to a confusion over terminological significance (e.g., when a feature happens to resemble a department title), on the other hand.

\subsubsection{Justifying and Questioning Hierarchies} 
\label{sec:hierachies}

A further paradigmatic explanation strategy was related explicitly to the explanation generated by the similarity technique, i.e., \textit{cluster regions}. We found that the discrete clusters led participants to search for specific hierarchies between clusters (e.g., P5 ``Maybe [the green cluster] is research results, and the [red cluster] is research?''). P2 noted that the green cluster seemed to be ''geological,'' indicating that a particular research department would most probably dominate this cluster. While participants used the cluster regions effectively to solve the tasks provided in phase I, much like the explanations generated from feature importance techniques, we also found that the search for hierarchies showed that participants had a distinctly contextual notion of similarity, which did not map to the probabilistic measure of similarity of the explanation method used. In this sense, the cluster regions were ineffective from a usability standpoint, given that they were not interpreted in a `technically correct' manner. However, seen more broadly, the various ways in which participants tried to make sense of the cluster regions also indicate that it provoked participants not to take the explanation offered as a given fact. Our finding on this explanation, and the underlying similarity technique, i.e., explanation method, are, therefore, more ambivalent than a focus on usability or technical correctness would suggest.

\subsection{Conceptual Explanation Strategies}

Conceptual explanation strategies are those in which a matter at hand calls for generating novel concepts or adopting those from other domains. Specifically we found in our co-design workshop that the \textit{uncertainty landscape}, though not the one used most overall, was applied by participants to interpret the data in two ways distinct from all other explanations. 

\subsubsection{Attribution of an `Algorithmic Perspective'} 
\label{sec:attribution}
We found that the uncertainty landscape led participants to attribute an \textit{algorithmic perspective} to this explanation in terms of its functional principle (e.g., P5: ``Here [the algorithm] is not so sure''). This was shown in participants' reasoning about the functional principle of the visualization, such as how the ``binding of [cluster] affiliation and [word] occurrence may result [in the particular landscape]'' (P3). Additionally, participants reflected on the significance of the relationship between words and uncertainty, noting that a project may be classified as uncertain despite ``matching [cluster and project] topwords'' (P1). Furthermore, when enacting the visualization in phase II, participants attributed the effects of algorithmic characteristics on data by way of the uncertainty landscape more explicitly. P1, for instance, tied it explicitly to the question of ``which [cluster] we would \textit{assign} <emphasis> [the new projects] to''. P5, additionally, observed uncertainty ``trenches'' between clusters where ``the machine doesn't know where to go.'' We argue that the attributions of an `algorithmic perspective' which allowed stakeholders to probe the ML-driven visualization and reflect on the algorithmic representation of their context more critically are particularly noteworthy.

\subsubsection{Questioning the Coherence of Explanations} 
\label{sec:coherence}
We found that both groups A and B reinterpreted the light regions of the topography (indicating a high cluster fit) to be indicative of \textit{exemplars}, meaning that the projects located in light regions were semantically `dominant' (e.g., P4 when comparing local instances to clusters: ``This light project is significant for the cluster, the rest follows''). Group A, especially, supplemented their strategies subsequent to this reinterpretation: uncertainty\textit{-as-exemplars} became a key conceptual explanation strategy for proposing and searching for explanations.
Interestingly, the similarity technique in combination with the uncertainty landscape, also became a way to reflect more directly on contextual hierarchies. P6, for example, stated that the uncertainty landscape seemed to ``correspond to the cluster regions.'' Similarly, P5 noted that the areas between the red and blue clusters were ``always uncertain.'' Additionally, P1 argued in phase II that the green cluster offers ``more [peaks] for comparison.'' 
This indicates that the similarity technique in combination with the uncertainty landscape offered opportunities for stakeholders to search for new hierarchies and potentially similarities among research projects in their context. This finding, furthermore, suggests that the combination of explanations with the uncertainty landscape allows non-ML experts to expand their basis for sense-making. Differently put, juxtaposing the discrete outputs of other techniques with the continuous uncertainty landscape helped our participants to reflect on the internal coherence of explanations.

\subsection{Presuppositional Explanation Strategies}

Presuppositional explanation strategies indicate the more general frames of reference that people hold regarding a context such as `their' experiences. We found two explicit over-arching themes as presuppositional explanation strategies especially based on participant statements and interactions in phases III and IV: \textit{organizational hierarchy} and \textit{internal} and \textit{external socio-material relations}. These themes allow us to understand the diverse ways in which stakeholders in our use case have become ``accultured''~\cite{reis_louis_acculturation_1990} to a particular view of the institution.

\subsubsection{Organizational Hierarchy as a Heuristic for Interpretation}
\label{sec:hierarchy}
We found that the contextual attributions participants expressed in relation to the explanations provided in the visualization explained specific attitudes about the organizational hierarchy of the research institution (e.g., P2: ``We don't do any [\textit{emphasis}] work with that research department''). The organizational structure was the obvious lens through which to view the data for group A. This became evident when, ten seconds into phase III, group A settled on building their representation of research projects (cf.~\autoref{fig:part2}, left) based on the organizational structure (P1: ``We're rebuilding the organizational diagram''). Group B followed the hierarchy of the research institution as well (cf.~\autoref{fig:part2}, center) though integrated references to the role of the public, noting the institution as a ``site of science encountering society'' (P3). Group C chose to distinguish research topics by separating past, ongoing and future research topics into intersecting planes (cf.~\autoref{fig:part2}, right). However, this group remarked that the organizational hierarchy was challenging, observing that it could have been represented as ``a black box'' (P5). This finding indicates that while the organizational hierarchy inevitably figures in specific explanation strategies, it does not lead to the same interpretations for all stakeholders. Therefore, our initial design intuition of using a color schema based on the organizational hierarchy for identifying research projects in the visualized data is problematized: this design choice may, in fact, obscure the reflective sense-making of data instead of supporting stakeholders.

\subsubsection{Socio-Material Relationships}
\label{sec:relations}
Participants stressed the importance of socio-material relationships within and outside of the research institution both in constructing their own as well as in interpreting other groups' physical representations. Group B, for instance, noted that an important feature was the ``system environment'' (P3) of museum visitors, knowledge transfer activities (e.g., workshops or citizen science projects), multilateral cooperations in research projects and commercial endeavors. Group A, on the other hand, focused extensively on the role of physical infrastructures within the research institution (e.g., collections, libraries and laboratories) and the distribution of the associated expertise and responsibilities; noting that, for example, one department has ``many independent projects [...] that don't really have any connection to [infrastructures]'' (P1). Group C did not refer explicitly to either internal or external socio-material relationships due to their focus on how the institution's research activities change over time. However, group C foregrounded how the hierarchical view neglects the complexity of ongoing socio-material relationships between the institution, its employees and its research topics by referring to the dynamic between the ``constant [institution] and flexible [research topics]'' (P5). This finding suggests that referring to socio-material relationships helps stakeholders to formulate explanation strategies that contextualize the data visualized. In combination with the paradigmatic and conceptual explanation strategies found and the insight that the organizational hierarchy is an ambivalent heuristic at best, this finding, furthermore, indicates that representing infrastructures or external groups in the visualization may be more suited to supporting the sense-making of the data.
 
\subsection{Synthesis of Findings}
We could identify all three types of explanation strategies within the transcriptions of our co-design workshop. Firstly, we found concrete, paradigmatic explanation strategies when participants solved tasks, however, we saw that the explanations generated from feature importance \smash{\encircle{\textbf{\texttt{A}}}} \smash{\encircle{\textbf{\texttt{C}}}} and similarity techniques \smash{\encircle{\textbf{\texttt{B}}}} were predominantly interpreted in a utilitarian manner. Secondly, we found conceptual explanation strategies, specifically when participants engage with the uncertainty landscape generated by our regions of error technique \smash{\encircle{\textbf{\texttt{D}}}}. In contrast to the paradigmatic explanation strategies, these seem promising in supporting stakeholders to contextualize the data visualized while still remaining aware of the latter's technical nature. Thirdly, we found presuppositional explanation strategies when participants made and interpreted their own representations of the use case context, which allowed us to understand more deeply that our participants hold highly differential interpretations of the research institution itself.
It is important to note that no individual level of explanation strategy would have highlighted the utility of explanation strategies without the other. Group A's acceptance of the similarity explanations \smash{\encircle{\textbf{\texttt{B}}}}, for instance, would seem valuable from a `usability' standpoint, but by discerning the organizational hierarchy as a presuppositional explanation strategy, we saw that this acceptance had little to do with our data representation and the explanations provided. By contrast, the discovery of conceptual explanation strategies indicates that the uncertainty landscape \smash{\encircle{\textbf{\texttt{D}}}} was relatively unencumbered by presuppositions, and, instead, reflects on the results of the data visualization. We argue that this emphasizes the strong interdependence of technical explanations with stakeholders' contextual sense-making practices. Given the in-depth, multilevel knowledge about such contextual sense-making practices in our use case which we have gained from conducting only one co-design workshop, we, therefore, argue that these findings provide an initial indication of the utility for explanation strategies as an empirical-analytical dimension for design research in ML interpretability.

\section{Discussion}

Analyzing for explanation strategies allowed us to understand more precisely how participants integrated the explanations provided that were generated by selected explanation methods in their contextual sense-making. Situating our work as complimentary to the established approaches of ethnomethodology in HCI and CSCW (cf.~\autoref{ssec:expl}), we, therefore, suggest that explanation strategies as an empirical-analytical lens offer a humble step towards a more specific participatory design agenda in the ML interpretability space. In the following, we firstly discuss the methodological implications of our research. We then highlight two specific design concerns for explanations for non-technical stakeholders. Lastly, we suggest that similar empirical-analytical lenses from theories of technological mediation are highly promising for future work in the interpretability space.


\subsection{Methodological Implications of Explanation Strategies}
We propose that explanation strategies allow participatory design researchers to probe with stakeholders what kind of contextual sense-making practices exist, how these are enacted with (novel or existing) technological artefacts and where opportunities for discussion, elaboration or intervention present themselves. Our co-design workshop, in this sense, was a proof-of-concept for this proposal from which we derive two principal contributions. Firstly, we propose that explanation strategies are a promising object of study for participatory design research in the ML interpretability space. Secondly, we sketch one possible future application of this proposal regarding explainability scenarios.

\subsubsection{Studying Technologically Mediated Explanation Strategies}

Dove and Falyard have argued that HCI research should not pursue unilateral interpretability concerns, such as transparency or explanations, but rather find ways to interrogate development assumptions on interpretability in collaboration with stakeholders~\cite{dove_monsters_2020}. Participatory design methods have been posited to fill this gap; however, how this can be done in an actual context of use has not yet been exhaustively elaborated (cf.~\autoref{ssec:pdML}). 
Equally, HCI and CSCW research drawing on ethnomethodology does not necessarily provide a specific enough analytical dimension for ML interpretability. 
We propose that explanation strategies can function as a heretofore missing object of study for participatory design research in the ML interpretability space, functioning as the `what' that links context, explanation method and explanations. People do not interpret ML technologies in isolation in actual application contexts but instead always within and against a diverse backdrop of existing themes and practices. We found that our framing of explanation strategies allowed us to connect both the lived experience of stakeholders and the necessary technical explanations of our ML-driven visualization system. In our use case, for example, the dominance of the presuppositional `organizational hierarchy' explanation strategy led to group A searching for hierarchies between clusters generated from the similarity technique. While the explanation was successfully used to solve tasks and could, therefore, be seen as `usable,' our insight shows that this does not automatically translate into interpreting the data visualized in a reflective way. Explanation strategies, therefore, augment recent proposals (cf.~\cite{wang_designing_2019}), reflecting not only how particular interpretability measures are required due to how humans think about their worlds, but also how the latter is shaped by technologies in conjunction with specific contextual sense-making practices. Furthermore, we also see a potential in explanation strategies to support critical reflection on the part of HCI and CSCW researchers. Exemplarily, if contextual explanation strategies contrast drastically with the assumptions of developers and designers, then explanation strategies, on the one hand, offer evidence for such contrast and, on the other hand, disclose what particular high-level conceptual divergences need to be urgently addressed. In sum, we propose that explanation strategies can constitute a meaningful object of study that can allow researchers to articulate more precisely what participatory design research methods deal with in the ML interpretability space and, thereby, enable more robust, theory-driven and reflective analysis of empirical observations.

\subsubsection{Co-Designing Explainability Scenarios}

Our application of explanation strategies was limited to investigating their utility for understanding interpretability concerns in an actual context of use. However, given our results, we also see a potentially fruitful connection to Wolf~\cite{wolf_explainability_2019} and Andres et al.'s~\cite{andres_scenario-based_2020} development of explainability scenarios, which address the gap between technical opacity and interpretation in actual contexts of use in the form of exemplary vignettes. We propose that explanation strategies offer opportunities for a participatory design version of explainability scenarios. Researchers could co-design explainability scenarios most directly with stakeholders set in the latter's everyday work experience. Explanation strategies can here serve as the object of study that allows researchers to analyze how various types of sense-making make such scenarios meaningful to stakeholders.
Additionally, explanation strategies could also be integrated both \textit{a priori} and/or \textit{a posteriori} in participatory design research processes of explainability scenarios. 
On the one hand, explanation strategies learned from a co-design workshop such as ours could be used to formulate more contextually informed explainability scenarios, which could then be discussed with another group of stakeholders to probe the validity of the explainability concerns discerned.
On the other hand, researchers could also design explainability scenarios as prompts or ``boundary objects''~\cite{star_institutional_1989} for co-design workshops with stakeholders to explain what explanation strategies actually exist \textit{in situ} and whether these diverge from researchers' assumptions and, if so, in what way. The potential to combine explanation strategies with explainability scenarios is, thus, a first indication of the potential of our concept to contribute to critical and reflective design research in the interpretability space (cf.~\cite{EhsanRiedl2020:HumanCenteredXAI}).


\subsection{Design Implications for Explanation Methods for Non-ML Experts}
Our study results also offer implications for the design of explanation methods for non-ML experts, which we discuss briefly below. Such findings are in line with Dourish's notion of ``appropriable technologies''~\cite{dourish_appropriation_2003} and, should, therefore be seen as promising for lowering the sense-making threshold of ML technologies for non-ML experts.

\subsubsection{Enable Combinations of Explanation Methods to Support Explanation Strategies}

We found that all groups generally combined the explanations provided extensively for diverse interpretations. We, therefore, assume that our ML-driven visualization system should support this combinatory approach to explanation methods. Thus, instead of a visual separation into panels (e.g., as in Hohman et al.'s prototype~\cite{hohman_gamut:_2019}), explanation interfaces should allow for combining, for example, (de-)selecting and stacking explanations \textit{together}.
Our insight is supported by Binns et al.~\cite{binns_its_2018}, who note that interpretations need a multitude of available ``explanation styles.'' Furthermore, Dodge et al.~\cite{dodge_explaining_2019} have argued that ``hybridizing'' explanation methods can allow people to understand different scopes of an ML pipeline, such as specific qualities of a model or individual data predictions. We add to this argument our finding that contextual explanation strategies offer concrete impulses and guidance for this design implication, such as combining our region of error technique (here \smash{\encircle{\textbf{\texttt{D}}}}) with the similarity technique (here \smash{\encircle{\textbf{\texttt{B}}}}) to counteract hierarchical interpretations. Similarly, our findings suggest that feature importance techniques (here \smash{\encircle{\textbf{\texttt{C}}}}) should be treated with care in actual application contexts: words have specific, contextual significance, unlike in NLP pipelines where words are essentially non-semantic vectors. Therefore, assumptions about algorithmic significance should be counterbalanced by an understanding of contextual explanation strategies.

In sum, we saw that stakeholders combined explanations, which is a productive insight concerning their usability. On a deeper level, however, we learned which explanation strategies exist and which explanations supported stakeholders to reflect on the data visualized. In this sense, explanation strategies can be useful in the design of explanation methods as they constrain what these both could and should provide. Given the vast amount of explanation methods and their resulting explanations, not to mention the XAI algorithms which can be used, we argue that this is another valuable insight that explanation strategies can offer.

\subsubsection{Supplement Explanation Methods with Contextual Cues}

We found that the output (explanations) of the explanation methods provided should be supplemented by a socio-material perspective on the existing context (e.g., existing infrastructures, internal and external relationships) to enable diverse stakeholders to combine or contrast our ML-driven visualization with their explanation strategies. This can be seen as an extension of the ``focus + context'' principle from the information visualization domain that involves displaying detailed information in combination with contextual information to support people's sense-making process~\cite{hauser2006generalizing}. In our use case, we saw a valuable integration between our visualized clusters (i.e., the thematic organized research projects) and the existing organizational infrastructural resources (in our case, physical collections and artefacts). The latter were used to qualify interpretations on the organizational hierarchy of the research institution by representing and interpreting their context, in a similar way to how the uncertainty landscape qualified other explanations. We, therefore, hypothesize that integrating infrastructural resources into the visualization design can, in combination with the explanations, further support stakeholders in reflecting on the data visualized. Explanation strategies can, therefore, also be valuable for the design of applications that make use of explanation methods: researchers can discover what particular contextual matters can serve as contextual cues (e.g., ``This reminds me of \textit{x}.'' or ``Maybe this is because of our \textit{x}.'') by analyzing sense-making through explanation strategies.


\subsection{Considering Technological Mediation in ML Interpretability Research}

As more and more ML interpretability research seeks insights from the humanities (cf.~\cite{miller_explanation_2017}) and participatory design research methods are called for in this space, we argue that insights from the philosophy of technology similar to Hubig's can contribute novel research trajectories to the ML interpretability space. Consequently, we propose that post-phenomenology can play a vital role similar to its adaptation in HCI design research (e.g.,~\cite{hauser_annotated_2018,benjamin_machine_2021}), which Frauenberger has referred to as ``entanglement HCI''~\cite{frauenberger_entanglement_2019}. 
Integrating post-phenomenological investigations into interpretability research can offer further empirical-analytical opportunities and refine and expand what we can address with the explanation strategies proposed. Post-phenomenology posits that technological artefacts \textit{mediate} the practical, everyday relations between humans and `their' world. The design of technological artefacts, thus, (inadvertently or purposely) leads people to engage in particular activities. Our analysis can be seen as an investigation in exactly this vein: how particular artefacts (in our case, explanations) ``invite or inhibit''~\cite{verbeek_what_2006} specific contextual sense-making practices. To this end, the schemata of human-technology relations proposed in post-phenomenology (cf.~\cite{ihde_technology_1990,verbeek_what_2006}) may further expand our understanding of explanation strategies. Our position as developers of explanations and designers, for example, implies that we have a specific \textit{hermeneutic relation} with the use case context: we `read' the latter as data through the components of our data analysis pipeline. The use case stakeholders also have a hermeneutic relation with the explanations, but it is configured differently as they already `live with' their context. And, depending on the use case, there may also be more radically divergent relations; for example, if our stakeholders had an \textit{embodiment relation} with a wearable device. Therefore, analyses of human-technology relations can further differentiate how explanation strategies, and the presuppositions and values they express, manifest differently across these distinct relations. In turn, such insights can offer designers clearer anticipatory guidance (cf.~\cite{verbeek_beyond_2015}) to understand how explanation strategies manifest through particular relations, and how these can be purposely addressed.

\section{Limitations}

Similar to participatory approaches in general, our co-design method relied on voluntary participants. Therefore, a certain ``invitational bias'' (cf.~\cite{lindstrom_politics_2016}) will inevitably affect the kind of explanation strategies enacted in the workshop. We hypothesize that future research will alleviate this, relying on a wider range of participants to compare the simulated `use-before-use' of the co-design workshop with the actual deployment of explanation methods. 
Moreover, the research presented is necessarily limited as it is fundamentally explorative. While we claim that explanation strategies can become a meaningful empirical-analytical dimension in the ML interpretability space, our use so far has been more preliminary. By only using explanation strategies in one use case so far, their robustness and external validity as an \textit{empirical}-analytical lens has not yet been evidenced. In further deployments of our approach and future work in HCI and CSCW research, and with, accordingly, more empirical data at the researchers' disposal, complementary and widely used methodologies, such as grounded theory (cf.~\cite{muller_grounded_2010}) or, indeed ethnomethodology (cf.~\autoref{ssec:expl}), could be used to, on the one hand, reflect on the theory-driven typology for explanation strategies presented here and, on the other hand, outline more granular dimensions for analysis. Additionally, from a more technical perspective, we only studied a limited range of explanation methods. Further explanation methods (e.g., example-based, counterfactual) may well provide further inclinations for particular explanation strategies, similar to how feature importance techniques were mostly taken up in a utilitarian manner by our stakeholders. 
However, though we acknowledge these limitations, we argue that we have laid out a promising basis for future research through our literature work, conceptual framing, detailed deployment account and extensive discussion. In this sense, our proof-of-concept application does not aim for completeness but instead encourages future research to investigate empirically whether general heuristics on the relationship between explanation methods and explanation strategies could be derived.

\section{Conclusion}

In this paper, we have proposed explanation strategies as a suitable empirical-analytical dimension for understanding how explanation methods and situated sense-making of ML systems come together in an actual use case. We motivated this work from a methodological gap in the interpretability space: while the design of explanations, explanation methods and their algorithmic realization have been studied widely, how these become connected with contextual sense-making practices has not been exhaustively elaborated. The purpose of explanation strategies is, therefore, to help one understanding how technical ML interpretability concerns connect to the kind of qualitative, context-sensitive methods, such as participatory design methods, and broader methodological strains of HCI and CSCW research, such as ethnomethodology. 
From our experience in framing, developing, deploying and analyzing the workshop, we are confident that explanation methods and ML interpretability more generally can and should be studied with participatory design methods. Accordingly, we argue that explanation strategies as an empirical-analytical lens offer a way forward to engage in participatory design research of ML interpretability contexts with non-ML experts.

\begin{acks}
This work is supported by the German Federal Ministry of Education and Research, (BMBF), grant 03IO1633 (``{IKON} -- Wissenstransferkonzept f\"{u}r Forschungsinhalte, {-methoden} und {-kompetenzen} in Forschungsmuseen''). We thank our participants for their enthusiasm and invaluable engagement, reviewers for their constructive feedback and Rony Ginosar, Arne Berger, Richmond Wong, Nick Merrill, Patricia Cornelio and Katrin Glinka for their tireless support in bringing this paper together.
\end{acks}

\bibliographystyle{ACM-Reference-Format}
\bibliography{group2022}


\end{document}